\documentclass[
 aip,
jcp,
 amsmath,amssymb,
 reprint,%
]{revtex4-1}

\usepackage{graphicx}
\usepackage{dcolumn}
\usepackage{bm}
\usepackage{color}
\begin{document}

\preprint{}

\title[Generalized Bond Order Parameters to Characterize Transient Crystals]{Generalized Bond Order Parameters to Characterize Transient Crystals}

\author{M. Isobe}
 \email{isobe@nitech.ac.jp}
\affiliation{Graduate School of Engeneering, Nagoya Institute of Technology, Nagoya, 466-8555, Japan}
\affiliation{Department of Chemistry, University of California, Berkeley, California 94720, USA}

\author{B.J. Alder}
 \email{alder1@llnl.gov}
\affiliation{Lawrence Livermore National Laboratory, P.O.Box 808, Livermore, CA 94551-9900, USA}


\begin{abstract}
Higher order parameters in the hard disk fluid are computed to investigate the number, the life time and size of transient crystal nuclei in the pre-freezing phase.
The methodology introduces further neighbor shells bond orientational order parameters and coarse-grains the correlation functions needed for the evaluation of the stress autocorrelation function for the viscosity.
We successfully reproduce results by the previous collision method for the pair orientational correlation function, but some two orders of magnitude faster.
This speed-up allows calculating the time dependent four body orientational correlation between two different pairs of particles as a function of their separation, needed to characterize the size of the transient crystals.
The result is that the slow decay of the stress autocorrelation function near freezing is due to a large number of rather small crystal nuclei lasting long enough to lead to the molasses tail.
\end{abstract}

\maketitle

\section{Introduction}

The slow dynamics of supercooled liquids and glasses have been actively discussed for thirty years from the view point of both spatial distributions of heterogeneities and long time correlations.
Recently, the role played by heterogeneities in the slow dynamics of glasses has been emphasized~\cite{berthier_2011} and characterized in terms of 4-point correlation functions.
The long time correlation appear in the slow decaying potential part of the shear-stress autocorrelation function (SACF) and has been called the ``molasses tail'' to differentiate it from the hydrodynamic origin of the long time tail in the velocity autocorrelation function~\cite{alder_1970a, alder_1970b, isobe_2008} and to emphasize its relation to the highly viscous glassy state~\cite{alder_1986}.
The decay of the SACFs have been investigated by mode coupling theory (MCT)~\cite{geotze_1992} and kinetic theory similar to the velocity autocorrelation function that applies to the kinetic part of the shear autocorrelation function, but not to the potential part, which is central to the molasses tail.
Numerical studies eventually proved that the long time tail of both the kinetic and potential parts have a power decay consistent with MCT, however, the amplitude of the SACF in dense fluids was found to be orders of magnitude greater than predicted.
This discrepancy was ascribed to the possibility that the numerical results were not sufficiently long to resolve the long time correlations.
However, the cause of this tail is likely due to the slow structural relaxation in the dense liquid around the peak of the structure factor rather than by hydrodynamic phenomena at long wave length.
The existence of transiently crystal nuclei was also demonstrated in colloid experiments~\cite{conrad_2006} and Langevin dynamics simulation in glassy dense system~\cite{kawasaki_2010}.
Transient ordering mechanism in a quasi-two dimensional liquid near freezing was also investigated by Sheu and Rice\cite{sheu_2008a,sheu_2008b}.
However, the microscopic mechanism of the stress field relaxation has not been thoroughly investigated~\cite{furukawa_2009, kim_2005, levashov_2011, abraham_2012}.

Twenty years ago, Ladd and Alder have speculated that the long time tail of the shear stress auto-correlation function near the solid-fluid transition point in the hard sphere system is due to transient crystal nuclei formation~\cite{ladd_1989}.
They found that the potential part of the SACF and the angular orientational auto-correlation function (OACF) are identical in the long time limit and show non-algebraic decay in time.
Since the evidence suggested that the reason for non-algebraic decay is structural relaxation rather than hydrodynamic flow, an attempt was made to understand this slow decaying mechanism by decomposing the OACFs into two-, three-, and four-body correlations, however, the four-body correlations have not been obtained accurately due to computer limitation.
The prediction of the cooling rate necessary to prevent crystallization requires knowledge of the rate of growth of a cluster the size of a critical solid nuclei and the time they exist in this transient state, and can only be obtained from the four body correlation.

In our previous work~\cite{isobe_2009, isobe_2010}, we have investigated the slow decay of the pair, $C_2$, orientational autocorrelation function in a two dimensional system consisting of elastic hard disks at a single density near the solid-fluid transition point, placed in a square box with periodic boundary conditions, using a modern fast algorithm based on event-driven Molecular Dynamics (MD) simulation~\cite{isobe_1999}.
The time evolution of various sized cluster were detected by using the bond orientational order parameter~\cite{halperin_1978,steinhardt_1983,lechner_2008}.
Near the fluid-solid phase transition, we found three regimes in the relaxation of the pair orientational autocorrelation function, namely the kinetic, molasses (stretched exponential), and diffusional power decay (pairs breaking apart).
We confirmed the non-algebraic decay (stretched exponential) at intermediate times presumably due to the existence of various sized solid clusters at high densities decaying at different rates.

Then, we focused on the rapidly increasing time with increasing density for the decay of the OACFs and were able to establish the length of time for which the biggest such nuclei exists at each density.
The largest cluster near the freezing density was found to be only a few sphere diameters in size and to persist for typical argon parameters for only about 30 picoseconds.
We also compared the results to theoretical predictions of the final power law decay~\cite{isobe_2010}.

To make further quantitative progress, we need to investigate the OACF of the quadruplet component, $C_4 (\Delta R,t)$, as a function of the distance between the two colliding pairs $\Delta R$.
From this information it will be possible to tell how the cluster size distribution changes with time and density, and, subsequently, determine how fast one has to increase the density to get a glass instead of a crystal.
Because this is computationally a very demanding task, we introduce two methodologies in this paper, which are more efficient methods for analysis of the quadruplet contribution to the orientational auto-correlation function.
One is a more efficient coarse grained algorithm for calculating pair and quadruplet contributions to the OACFs, rather than the previous collision based calculation.
The other is the extension of the usual bond orientational parameter $\phi_6^i$ to a higher order one involving further neighbor shells.
We demonstrate that the coarse grained results are in quite good agreement with the previous one, but two order of magnitude faster, allowing for the faster evaluation of the 4-body autocorrelation functions $C_4 (\Delta R,t)$.

This paper is organized as follows.
In Sec.~II, we describe how to detect further nearest neighbors and summarize details of the new algorithm.
In Sec.~III, the results of the new improvement are described.
Finally, in Sec.IV, we summarize the results and discuss the relaxation time of transient clusters and their size distribution.

\section{Determination of Transient Crystals in Dense Liquids}

In this section, we explain how to categorize neighboring shells.
Then, generalization of the bond orientational order parameter is described in order to calculate the autocorrelation of the orientational function.

\subsection{Detecting Higher Nearest Neighbors}

\begin{figure}
\center
\includegraphics[scale=0.45]{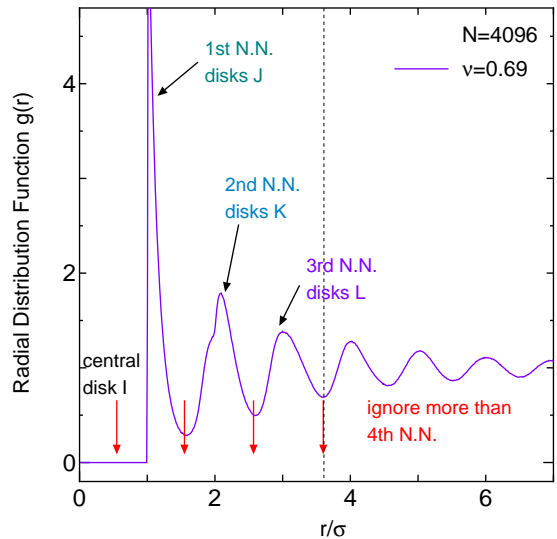}
\caption{
Radial Distribution Function at $\nu=0.69$ with the cut-off arrows.
}
\label{rdf069}
\end{figure}

To consider further neighbors than nearest neighbors systematically, we define neighbor shells based on the minima of the radial distribution functions (RDF) for each packing fraction $\nu$, which are obtained by an independent calculations via event-driven MD.
The radial distribution function, $g(r)$, at $\nu=0.69$ in a two-dimensional (2D) system composed of $N=4096$ hard disks with a diameter $\sigma$ are shown in Fig.~\ref{rdf069}.
We show 4 red arrows for 3 shell radii beyond the central particle, the 1st nearest neighbors (N.N.), 2nd N.N., and 3rd N.N., which are named by the shell index $I$, $J$, $K$, and $L$, respectively.
(e.g., $J$ indicates the particles belong to 1st N.N. against the central particle $I$.)
Note that the 2nd N.N. peak has a shoulder nearby, which likely indicates that the transient crystal becomes significant in dense liquids.
The result is the shell radii given as the cut-off distance $r_{cut}$ for each packing fraction, as summarized in Table \ref{tbl:1}.
We only consider 3rd N.N. shell in this paper, however, considering further neighbors would be straightforward, but such large clusters were found to be rare.

\begin{table}
\begin{center}
\begin{tabular}{cccc} \hline \hline
 $\nu$  & 1st N.N. & 2nd N.N. & 3rd N.N.  \\ \hline
$0.69$ & $1.579$ & $2.592$ & $3.606$  \\ 
$0.65$ & $1.616$ & $2.638$ & $3.660$  \\ 
$0.57$ & $1.702$ & $2.747$ & $3.838$  \\ 
\hline \hline
\end{tabular}
\end{center}
\caption{
The distances from the central particle $i$ to further neighbor are shown in units of $\sigma$ at various packing fraction.
}
\label{tbl:1}
\end{table}

\begin{figure}
\center
\includegraphics[scale=0.30]{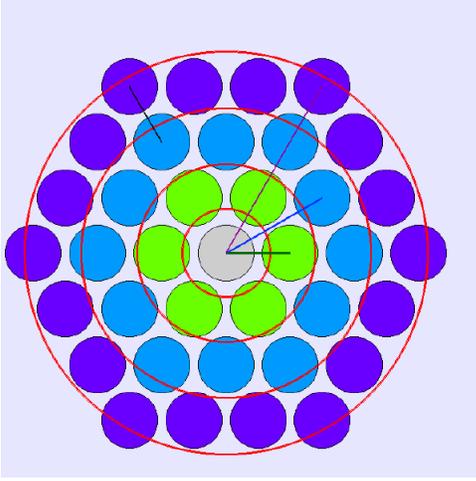}
\caption{
Perfect crystal configuration and neighbor shells at $\nu=0.69$.
}
\label{perfect069}
\end{figure}

\begin{figure}
\center
\includegraphics[scale=0.45]{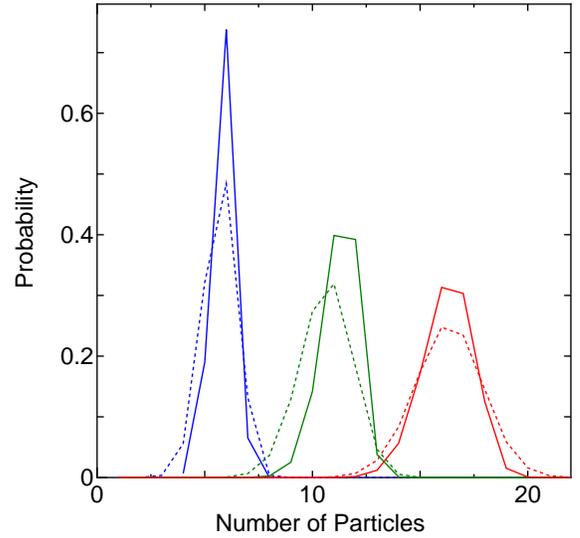}
\caption{
Probability distributions of particle number for each neighbor shell at $\nu=0.69$ (---) and $\nu=0.57$ ($\cdots$) are shown.
}
\label{PDF_NN_NUM}
\end{figure}

To justify the concept of the above categorization of neighbors, we consider the perfect crystal configuration for $\nu=0.69$ as a reference, which is shown in Fig.~\ref{perfect069}.
We also show $4$ red shells (circles) correspond to the cut-off radius in Fig.~\ref{perfect069}.
\begin{table}
\begin{center}
\begin{tabular}{ccccc} \hline \hline
$\nu$   & 1st (J) & 2nd (K) & 3rd (L) \\ \hline
perfect crystal  & 6 & 12 & 18 \\ 
$0.69$ & $5.93$ & $11.57$ & $17.23$ \\ 
$0.65$ & $5.88$ & $11.29$ & $16.71$ \\ 
$0.57$ & $5.68$ & $10.62$ & $16.32$ \\ 
\hline \hline
\end{tabular}
\end{center}
\caption{
The 1st to 3rd N.N. particle numbers for a perfect crystal and the actual particle number for various packing fractions.
}
\label{tbl:2}
\end{table}
The simulation results on the probability distribution for the particle number of neighbors and its averages are shown in Fig.~\ref{PDF_NN_NUM} and Table \ref{tbl:2}, respectively.
We hereby justify that neighbors can be detected for 1st to 3rd N.N. by setting simple cut-off distances from the central particle based on the RDFs.

\subsection{Decomposition of the Orientational Factors and Autocorrelation Functions}

To investigate the temporal properties of clusters in the liquid state, the time correlation functions of dynamical variables $\sum_{i,j} {\cal A}({\bf r}_{ij},t)$ are considered~\cite{balucani_1994}, where ${\bf r}_{ij}$ in the relative distance between the position of two particles ${\bf r}_i$ and ${\bf r}_j$.
The time correlation function can be written as,

\begin{equation}
C(t)=\left< \sum_{\substack{i,j \\ (i \neq j)}} {\cal A}({\bf r}_{ij}(0)) \sum_{\substack{k,l \\ (k \neq l)}} {\cal A}({\bf r}_{kl}(t)) \right>.
\end{equation}
We have investigated the potential part of the SACF $\langle J^P_{xy}(t)J^P_{xy}(0) \rangle$ relevant for the molasses tail, where $J^P_{xy}$ is the potential part of the momentum current $J^P_{xy}$.
For a pair-wise potential $\phi(r_{ij})$, $J^P_{xy}$ is

\begin{equation}
{\cal A}({\bf r}_{ij})=J^P_{xy}=\frac{1}{2} \left(\frac{n m^2}{Nk_BT}\right)^{\frac{1}{2}} \frac{x_{ij}y_{ij}}{r_{ij}} \phi'(r_{ij}),
\end{equation}
\noindent
where $n$ and $m$ are number density and mass of disks, $k_B$ and $T$ are Boltzmann constant and temperature, $(x_{ij},y_{ij})=(x_i-x_j,y_i-y_j)$ are the relative positions between particles $i$ and $j$.
In the case of hard disks, this becomes,
\begin{equation}
J^P_{xy}(t)=\sum_\gamma -mb_{ij}\frac{x_{ij}y_{ij}}{\sigma^2}\delta{(t-t_\gamma)}.
\label{eqn:p_ss}
\end{equation}
where $b_{ij}={\bf v}_{ij}\cdot{\bf r}_{ij}$ and $\sum_\gamma$ means the accumulation of collisional contributions at the colliding time $t_\gamma$.
In general, $C(t)$ can be decomposed into pair $C_2(t)$ ($ij-ij$ pair), triplet $C_3(t)$ ($ij-ik$ pair), and quadruplet $C_4(t)$ ($ij-kl$ pair) contributions~\cite{balucani_1994, ladd_1979},
\begin{eqnarray}
C(t)& =& \left< \sum_{\substack{i,j \\ (i \neq j)}} {\cal A}({\bf r}_{ij}(0)) \sum_{\substack{k,l \\ (k \neq l)}} {\cal A}({\bf r}_{kl}(t)) \right> \\
& = & 2 \sum_{\substack{i,j \\ (i \neq j)}} \left< {\cal A}({\bf r}_{ij}(0)) {\cal A}({\bf r}_{ij}(t)) \right> \nonumber \\
& &+ 4 \sum_{\substack{i,j,k \\ (i \neq j \neq k)}} \left< {\cal A}({\bf r}_{ij}(0)) {\cal A}({\bf r}_{ik}(t)) \right> \nonumber \\
& &+ \sum_{\substack{i,j,k,l \\ (i \neq j \neq k \neq l)}} \left< {\cal A}({\bf r}_{ij}(0)) {\cal A}({\bf r}_{kl}(t)) \right> \nonumber \\
& = & C_2(t)+C_3(t)+C_4(t).
\end{eqnarray}

Since velocities and positions are no longer correlated beyond a few mean collision times, only the orientational part of the SACF, namely the orientational autocorrelation function (OACF), $\langle O_{xy}(t)O_{xy}(0) \rangle$, needs to be studied~\cite{ladd_1989,isobe_2009,isobe_2010}.
$O_{xy}(t)$ is defined as 

\begin{equation}
O_{xy}(t)=\sum_\gamma \frac{x_{ij}y_{ij}}{\sigma^2}\delta{(t-t_\gamma)}.
\label{eqn:o_factor}
\end{equation}

\noindent
To avoid the delta function singularity of $O_{xy}(t)$ for hard particles, the alternative Einstein-Helfand expression~\cite{helfand_1960, alder_1970b} involving the second derivative, obtained by the numerical differentiation, is needed for calculating the correlation function,

\begin{eqnarray}
C(t) & = & \langle O_{xy}(t)O_{xy}(0) \rangle \nonumber \\
 & = & \frac{1}{2}\frac{d^2}{dt^2}\langle (G(t)-G(0))^2 \rangle,
\label{eqn:acf}
\end{eqnarray}

\noindent
where

\begin{equation}
G(t)=\sum_\gamma \frac{x_{ij}y_{ij}}{\sigma^2} \Theta (t-t_\gamma),
\label{eqn:acf2}
\end{equation}
\noindent
and where $\Theta (t)$ is the unit step function.
Note that there are three independent orientational factors ($O_{xy}$, $O_{yz}$, and $O_{zx}$) in 3D.
The pair and quadruplet contributions of OACF are defined as,

\begin{eqnarray}
C_2 (t) & \sim & \frac{1}{2}\frac{d^2}{dt^2} \left\langle \sum_{\substack{i,j \\ (i\neq j)}}^N G^{ij}(t)^2 \right\rangle, \label{eqn:C2} \\
C_4 (t) & \sim & \frac{1}{2}\frac{d^2}{dt^2} \left\langle \sum_{\substack{i,j,k,l \\ (i\neq j \neq k \neq l)}}^N G^{ij}(t) G^{kl}(t) \right\rangle, \label{eqn:C4}
\end{eqnarray}
\noindent
since $\langle G^{ij}(0)\rangle = \langle G^{kl}(0) \rangle =0$ ($\dot{G}^{ij}(t)=O^{ij}_{xy}(t),\dot{G}^{kl}(t)=O^{kl}_{xy}(t)$).
To ease calculating $C_2$ and $C_4$, we introduce a ``collision pair index'' 
\begin{equation}
\gamma_k=(\gamma_i-1)N-\gamma_i(\gamma_i-1)/2+\gamma_j-\gamma_i ,
\label{eqn:collision_pi}
\end{equation}
where $\gamma_i$ and $\gamma_j$ are particle indexes of colliding pairs, which identifies a given pair quickly, thus avoiding having to check whether the same collision pair has collided before.
For example, in case of $N=4$, the total number of 
collision pairs are $N(N-1)/2=6$, which can be listed as $(\gamma_i,\gamma_j)=(1,2),(1,3),(1,4),(2,3),(2,4),(3,4)$, where $\gamma_i < \gamma_j$.
By using the ``collision pair index'', we obtain $\gamma_k=1,2,3,4,5,6$ for each collision pair, respectively.
Therefore, it is convenient to deal with the collision pair as the sequential number to sort and insert into the array of correlation pairs of $G^{ij}(t)$.
This speeds up the calculation considerably.
All properties in this paper are normalized and indicated by $C^*(t^*)$, where $t^*$ is the reduced time $t^* = t/t_0$($t_0$ is the mean free time).

\subsection{Higher Order Orientational Factor Based on Course Graining Method}

Here, we propose an alternative methodology for calculating the OACF efficiently, which differs from the previous method based on collision events.
The new method needs only configurational data of particles at a certain discrete time $t=\lambda \times \Delta t$ instead of collisions, where $\lambda$ is an integer number and $\Delta t$ is taken at an interval e.g. the mean free time, $\Delta t=t_0$.
The essential idea of our improvement is the coarse graining (CG) of collision events in neighbor shells at discrete small times, in which we recognize the tagged particles as the candidates of collisions within $\Delta t$.
Therefore, we can calculate orientational factors by summing those candidates with the following slight modification,

\begin{eqnarray}
O_{xy}(t) & = & \sum_{I,J} \frac{x_{IJ}(t) y_{IJ}(t)}{(\sigma + \delta r)^2} \nonumber \\
& =& \sum_{I,J} \cos{\theta_{IJ}}\sin{\theta_{IJ}}, \label{eqn:CG_o_factor}
\end{eqnarray}

\noindent
where $\delta r = \sqrt{x_{IJ}(t)^2+y_{IJ}(t)^2}-\sigma$ ($\ll \sigma$) is the small gap distance between particle $I$ and $J$ and $\theta_{IJ}$ is the angle of vector ${\bf r}_{IJ}$ against a reference axis (e.g. $x$-axis).
Such a modification has a great advantage in efficiency since we do not wait for the actual particle collisions as in the event-driven scheme.
We restrict the particle pairs considered as the reference pairs at the start of each simulation as nearest neighbors.
This is expected to improve sampling drastically by quick sorting over choosing arbitrary reference particle pairs as we have done in the previous method~\cite{isobe_2010}.

\subsection{Generalized Order Parameter Based on Crystal Structure}

\subsubsection{Bond Orientational Order Parameter}

The usual bond orientational order parameter $\phi_6$ for a hard disk $i$ is defined by~\cite{halperin_1978},

\begin{equation}
\phi_6^i = \frac{1}{N_i} \sum_{j=1}^{N_i} \exp{(6{\rm i}\theta_i^j)},
\label{eqn:phi6}
\end{equation}

\noindent
where $N_i$ is the number of the nearest neighbors around the tagged particles $i$, and $\theta_i^j$ is the angle between the position vector from the disk $j$ to $i$ and an arbitrary fixed reference axis (e.g., $x$-axis).

Previously, two disks are defined as nearest neighbors if the separation is within a distance, say $1.4$ to $1.7 \sigma$.
This choice is reasonable from the view point of our definition by the RDFs.
The absolute value $\Phi_6^i=\sqrt{{\phi_6^i}^* \phi_6^i}$ takes on values between $0$ and $1$, and measures the degree of crystallization in terms of considering only nearest neighbors, where ${\phi_6^i}^*$ is the complex conjugate of $\phi_6^i$.
\begin{figure*}
\begin{minipage}{0.48\hsize}
\begin{center}
\includegraphics[scale=0.49]{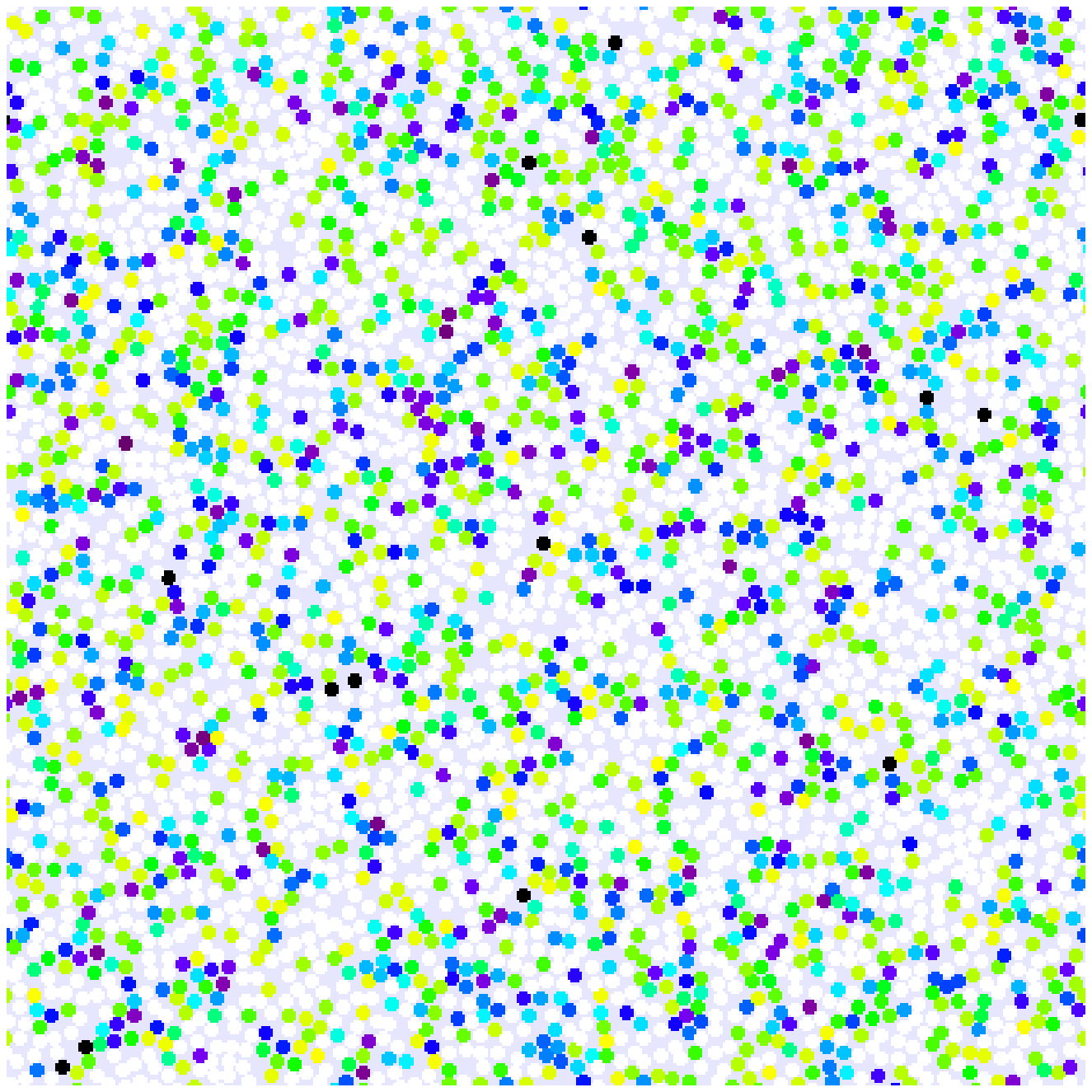}
\end{center}
\begin{center}
0.5 \includegraphics[scale=0.13]{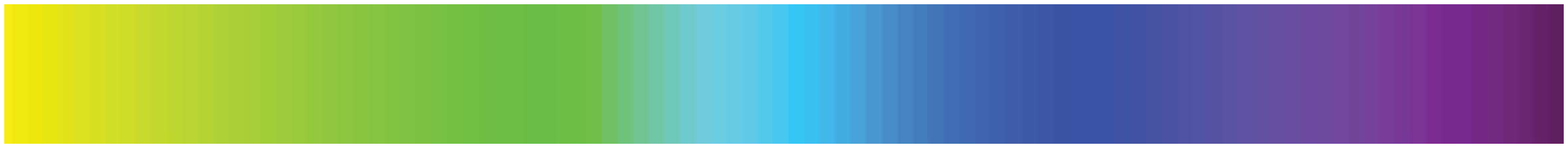} 1
\end{center}
\end{minipage}
\begin{minipage}{0.47\hsize}
\begin{center}
\includegraphics[scale=0.54]{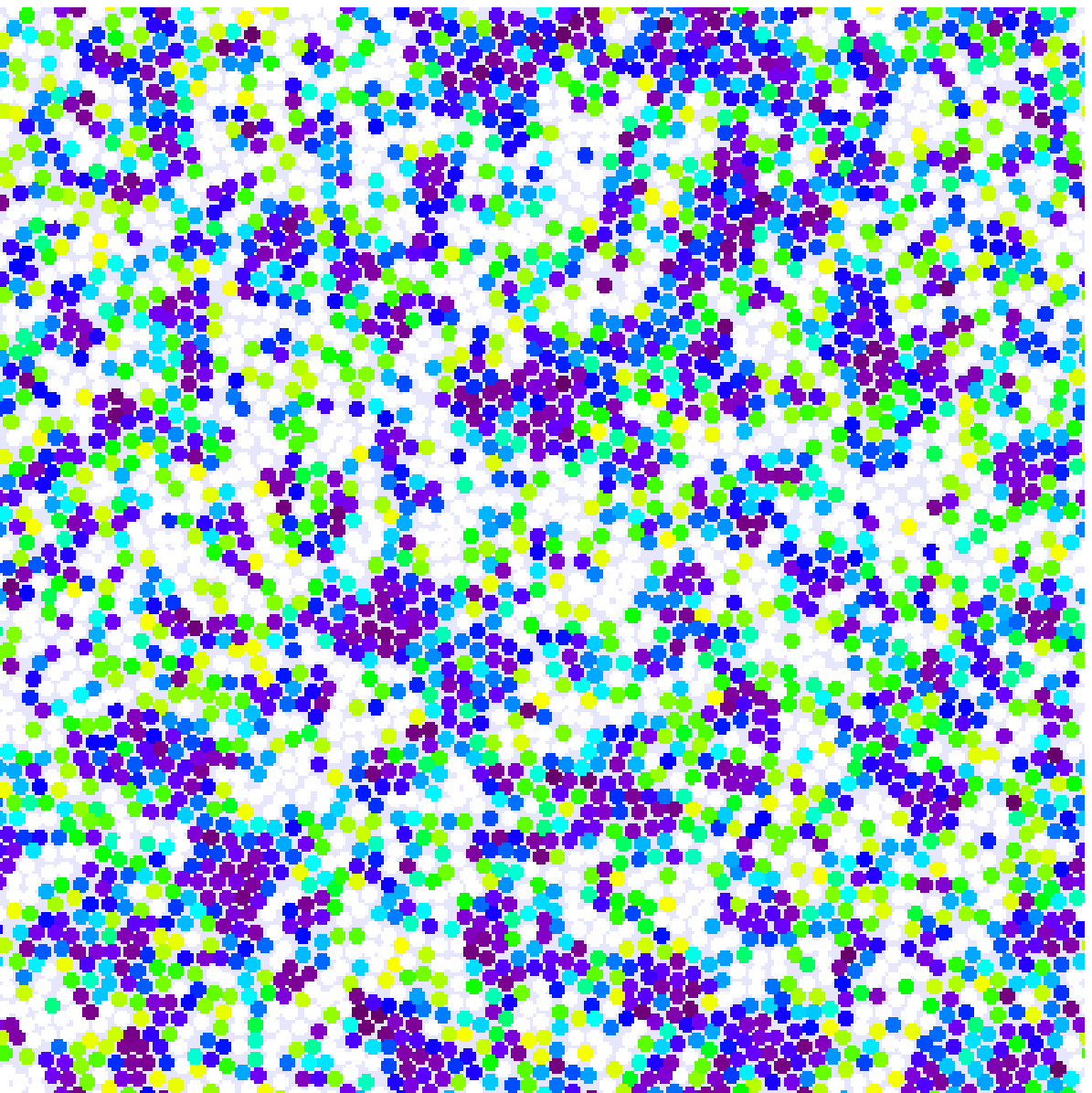}
\end{center}
\begin{center}
0.5 \includegraphics[scale=0.13]{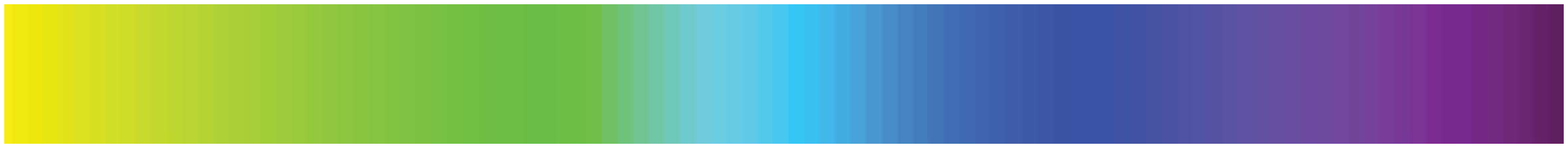} 1
\end{center}
\end{minipage}
\caption{The spatial distribution of $\Phi_6^i$ for $4096$ particles system at a given time for two packing fractions, $\nu=0.57$(left) and $0.69$(right).
The darker the region, the closer $\Phi_6^i$ is to unity.}
\label{fig:phi6_snapshot}
\end{figure*}
Figure~\ref{fig:phi6_snapshot} shows typical snapshots of the spatial distribution of $\Phi_6^i$ at the packing fraction $\nu=0.65$ (left) and $0.69$ (right), respectively.
The gradation in shading of the particles indicates the value of $\Phi_6^i$; the darker, the closer to unity.
We clearly observe the dramatic growth of several solid nuclei as the density nears solidification.
However, those solid nuclei will disappear after a certain transient time\cite{isobe_2010}.
To investigate those transient clusters more quantitatively, we extended the usual bond orientational order parameter $\phi_6$ toward further neighbors, that is, the 2nd to 3rd N.N. and investigate the relaxation time via autocorrelation functions.

\subsubsection{Angle for Neighbors based on Triangle Crystal Lattice}

Here, the angle between bond vectors for the generalized order parameter is considered.
It is easy to deal with the neighbors when we introduce the integer index $(\zeta, \eta)$ summarized in Table~\ref{tbl:4}.
If the particle is located on the crystal lattice with the side distance $a$, the position vector of particle ${\bf c}$ can be described as 

\begin{equation}
{\bf c}=\zeta{\bf a}+\eta{\bf b},
\end{equation}

\noindent
where $(\zeta, \eta)$ are integers and ${\bf a}=(1,0)a$, ${\bf b}=(\frac{1}{2},\frac{\sqrt{3}}{2})a$ are unit vectors of the crystal lattice.

\begin{table}
\begin{center}
\begin{tabular}{cc} \hline \hline
central particle (I) & (0,0) \\ \hline
1st N.N. (J) & (1,0), (0,1), (-1,1), (-1,0),(0,-1), (1,-1) \\ \hline
2nd N.N. (K) & (1,1), (-1,2), (-2,1), (-1,-1), (1,-2), (2,-1) \\ 
             & (2,0), (0,2), (-2,2), (-2,0), (0,-2), (2,-2) \\ \hline
3nd N.N. (L) & (3,0), (2,1), (1,2), (0,3), (-1,3), (-2,3) \\
             & (-3,3), (-3,2), (-3,1), (-3,0), (-2,-1), (-1,-2) \\
             & (0,-3), (1,-3), (2,-3), (3,-3), (3,-2), (3,-1) \\ 
\hline \hline
\end{tabular}
\end{center}
\caption{
$(\zeta, \eta)$ pairs for neighbors.
}
\label{tbl:4}
\end{table}

We next consider the angle of bond vectors.
If the particle positions are in a perfect crystal, that is ${\bf r}={\bf c}$, the bond vector between central particle $I$ and 1st N.N. $J$ can be defined as,

\begin{equation}
{\bf c}_{JI}={\bf c}_J-{\bf c}_I.
\end{equation}

\noindent
We can define $6$ kind of bond vectors (${\bf c}_{JI}$, ${\bf c}_{KI}$, ${\bf c}_{LI}$ ${\bf c}_{KJ}$, ${\bf c}_{LJ}$, and ${\bf c}_{LK}$) within 3rd N.N.
Angles $\theta$ between bond vectors for ${\bf c}_{JI}$ and ${\bf c}_{J'I}$ ($J'\neq J$) are easily calculated as $\theta = n' \times \pi/3 $, where $n'$ is an integer, which reduces to the usual calculation for $\phi_6$.
In general, the angle between bond vectors can be calculated by,

\begin{equation}
\theta_I^{JJ'}=\cos^{-1}{\left(\frac{{\bf c}_{JI}\cdot{\bf c}_{J'I}}{|{\bf c}_{JI}||{\bf c}_{J'I}|}\right)}
\end{equation}

We can define $432$ pairs of bond vectors within 3rd N.N., of which $6$ are for ${\bf c}_{JI}$, $12$ for ${\bf c}_{KI}$, $18$ for ${\bf c}_{LI}$, $72$ for ${\bf c}_{KJ}$, $108$ for ${\bf c}_{LJ}$, and $216$ for ${\bf c}_{LK}$.
We don't consider  ${\bf c}_{KJ}, {\bf c}_{LJ}$ in this paper.
Based on the perfect crystal, we calculate the angle probability distribution between the bond vector pairs (i) (${\bf c}_{JI}$, ${\bf c}_{JI}$) (${\bf c}_{KI}$, ${\bf c}_{KI}$) (${\bf c}_{LI}$, ${\bf c}_{LI}$), (ii) (${\bf c}_{JI}$, ${\bf c}_{KL}$).
Thus, the concept of bond orientational order parameter for further neighbors can be extended.

\subsubsection{Generalized Order Parameter Based on Crystal Structure}

The usual $\phi_6$ order parameter is obtained from the angle between the position vector from disk $J$ to $I$ and ``an arbitrary fixed reference axis''.
To consider the angle between 1st and 2nd neighbors pairs (and more), the bond angle can be redefined by relative vectors between $I$-$J$ and $I$-$K$ (etc..), instead of ``an arbitrary fixed reference axis''.
The complex generalized order parameter for a tagged particle $I$ with the actual position vectors ${\bf r}_{JI}$ can be generalized by

\begin{eqnarray}
\phi_s^I & = & \frac{1}{N_I(N_I-1)} \sum_{J<J'}^{N_I} \chi_s ({\bf r}_{JI},{\bf r}_{J'I}) \label{eqn:GOP-1}, \\
    \chi_s({\bf r}_{JI},{\bf r}_{J'I}) & = & \exp{({\rm i} s \theta({\bf r}_{JI}, {\bf r}_{J'I}))}, \label{eqn:GOP-2} \\
    \theta({\bf r}_{JI},{\bf r}_{J'I}) & = & \theta_I^{JJ'} = (\theta_I^{J'} - \theta_I^J) = \cos^{-1}{\left(\frac{{\bf r}_{JI}\cdot{\bf r}_{J'I}}{|{\bf r}_{JI}||{\bf r}_{J'I}|}\right)}.
\label{eqn:GOP-3}
\end{eqnarray}

\noindent
If $s=6$ and ${\bf r}_{J'I}=(1,0)$ (i.e., unit vector of $x$-axis) are fixed, we can reduce the above expression to the usual $\phi_6$ order parameter described in eq.~(\ref{eqn:phi6}).

\section{Results}

The systems considered are hard disks placed in a $L_x \times L_y$($=A$) square box with periodic boundary conditions.
Initially, the simulation systems for each packing fraction $\nu$ $(=N\pi (\sigma/2)^2/A$) are prepared in an equilibrium state by a sufficiently long preliminary run.
For the close packed area $A_0$, $\nu=\pi/(2\sqrt{3}(A/A_0))$.
The system evolves through collisions, using an algorithm based on event-driven MD simulation~\cite{isobe_1999}.
Most of the calculations are done with a relatively small particle numbers of particles, $N=4,096$, since long time runs for calculating accurate tails of autocorrelation functions are needed. 
We confirmed that periodic boundary effects on the OACF don't appear for this system size~\cite{isobe_2010}.
The density is set at relatively dense values near the solid-fluid transition point $\nu_c$ near $0.70$, namely at $\nu = 0.69, 0.65$, and $0.57$, primarily to compare with the previous results~\cite{isobe_2009,isobe_2010}.


\subsection{Total and Pair Orientational Autocorrelation Functions}

The algorithm for $C_2(t)$ is described in the following steps,

\begin{enumerate}

\item Prepare the list of vectors for particle pairs within the 3rd N.N. shells at $t=0$, which are likely collision candidates in a short time.
This is called ``a reference particle pair list''.
This list is similar to the neighbor list known in the standard MD technique to increase efficiency.
Note that if we only register particles within 1 N.N. shell, they can escape from that shell after a short time and therefore, the 3rd N.N. choice is a better one.

\item Sort the above list vectors by sequential number according to the collision pair index $\gamma_k$ (eq.~(\ref{eqn:collision_pi})), where $\gamma_i$ and $\gamma_j$ are the index of the colliding candidate particles.

\item During the event-driven MD simulation at each discrete time $t=\lambda \Delta t$ ($\lambda=1,2,3,\cdots$), we list particle pairs within 3rd N.N. shell into such arrays and sort them sequentially by using the pair list index.

\item Next, compare the pair particle index $\gamma_k$ obtained from configurations at time $t$ with the list of reference pair index prepared at time $t=0$.

\item If the pair particle index is found to be the same as the reference pair index for the 1st N.N., we insert it in the CG modified orientational factor $O^{ij}_{xy}(t)$ (eq.~(\ref{eqn:CG_o_factor})).
If the particle index is not the same as in the list of reference pairs, we discard it.

\item After the simulation is performed for a long time we obtain averages for $\gamma_k$ and hence $C_2(t)$ via the Einstein-Helfand formula with time resolution $\Delta t$.

\end{enumerate}

\begin{figure}
\center
\includegraphics[scale=0.45]{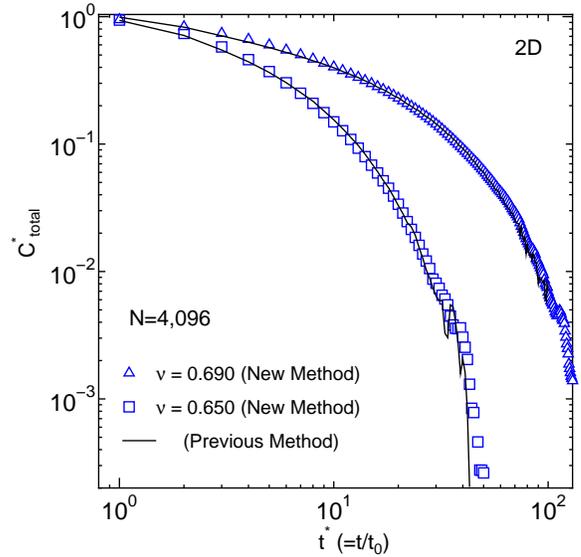}
\caption{
Comparison on the OACF ($C_{total}$) at two densities between previous collision-based method and new method. The vertical axis is normalized and the horizontal axis is the scaled time by the mean free time, $t_0$.
}
\label{fig:2d_oacf_cg}
\end{figure}

\begin{figure}
\center
\includegraphics[scale=0.45]{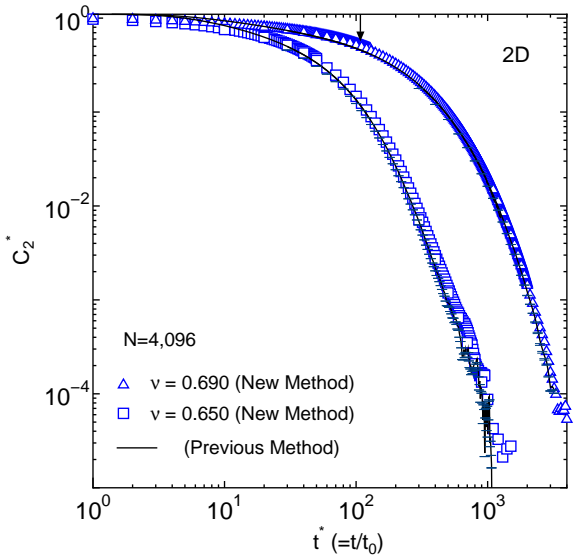}
\caption{
Comparison of $C_2$ at two densities between previous collision-based method and new method. The vertical axis is normalized and the horizontal axis is the scaled time by the mean free time, $t_0$.
}
\label{fig:2d_c2_cg}
\end{figure}

In Figs~\ref{fig:2d_oacf_cg} and \ref{fig:2d_c2_cg}, comparison of the $C_{total}$ and $C_2$ at two densities between the previous collision-based method and the new method are shown.
The relaxation time and long time behavior of both $C_{total}$ and $C_2$ obtained by the new method are in fairly good agreement with that of the previous method at both densities~\cite{isobe_2010}.
However, the efficiency proved to be drastically improved.
Although the efficiency depends on the particle number $N$ which determines the total array for particle pairs, the efficiency was more than $16$ and $77$ times faster than previously for the total and pair autocorrelation function at $(N, \nu)=(4096,0.69)$, respectively.

\subsection{Quadruplet Orientational Autocorrelation Functions}

We then showed that the new method can calculate higher order correlations such as the distance dependence of the autocorrelation for the quadruplet contribution, which cannot be resolved by the collision based calculation~\cite{isobe_2010}.

The $C_4$ algorithm is basically constructed as was $C_2$, however, it is more complex since we need to use more sorting procedures and searching method for detecting valid pairs for $i-j$ and $k-l$ within 3rd N.N.
The main computational task is to search the particle pair index in the reference particle pair list for both $i-j$ and $k-l$.
To detect the valid particle pairs for $i-j$ and $k-l$ within 3rd N.N. separated by a calculable distance at each discrete time is relatively easy, since those candidate pairs have already been registered at $t=0$.
Therefore, it leads to a significant improvement for calculating the $C_4$ autocorrelation functions.
Note that since not only $i-j$ pair but also $k-l$ must be the collisional candidates, $k-l$ pairs are searched under the condition that $k-l$ are also 1 N.N.

\begin{figure}
\center
\includegraphics[scale=0.45]{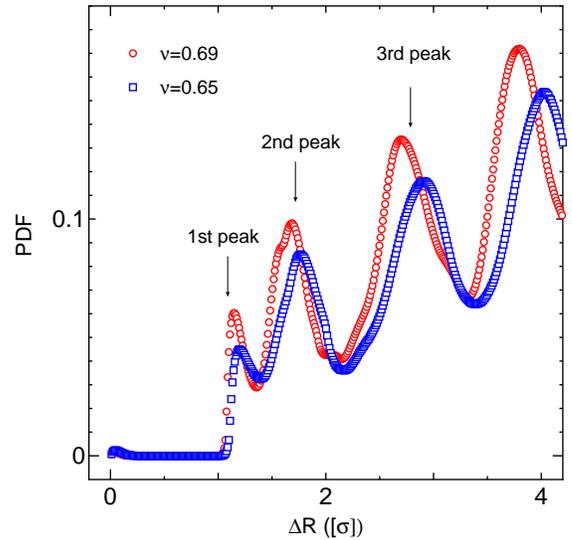}
\caption{
The probability distribution function of $\Delta R$ in the unit of $\sigma$ for $\nu=0.65$ and $0.69$.
}
\label{fig:ddr_hist}
\end{figure}

\begin{figure}
\center
\includegraphics[scale=0.45]{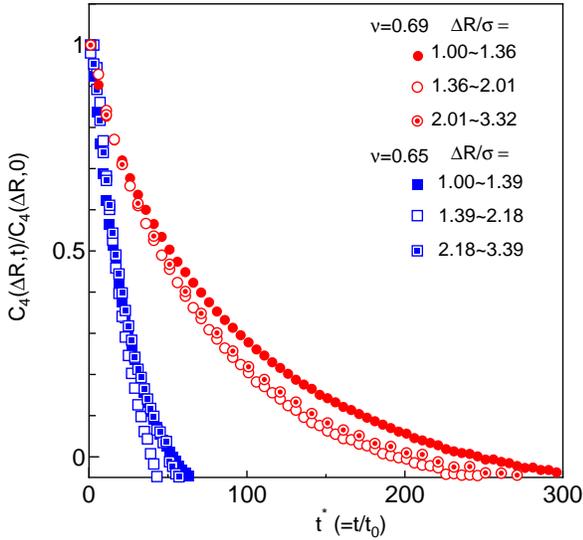} \\
\caption{
The distance dependence normalized $C_4(\Delta R,t)$ for each packing fraction $\nu=0.69$ and $0.65$ are shown.
}
\label{fig:c4r0690}
\end{figure}

Figure \ref{fig:ddr_hist} shows the probability distribution function of $\Delta R$ in units of $\sigma$, where $\Delta R$ is the separation between the center of mass of the particle pairs $i-j$ and $k-l$.
The contribution beyond the 4th N.N. are not calculated.
For a perfect crystal, the number of distinct quadruplet pairs ($i-j,k-l$) for tagged particle $i$ are $\sim 118.5 N$ in the $N$ particle system.
The actual sampling of quadruplet pairs are about $91$ pairs for each tagged particle $i$ at $\nu=0.69$.
Figure~\ref{fig:c4r0690} shows the time dependence of $C_4(\Delta R,t)$ normalized by $C_4(\Delta R, 0)$ for two packing fraction $\nu=0.69$ and $0.65$ for the $\Delta R/\sigma$ corresponding to 1st, 2nd, and 3rd peaks in Fig.~\ref{fig:ddr_hist}, respectively.

In $C_4(\Delta R, t)$ beyond $\Delta R/\sigma \sim 3.4$, the contribution of outer particles of 4rd N.N. shell gradually becomes dominant.
Computational costs are just a few times larger than for the $C_2$.
In Fig.~\ref{fig:c4r0690}, we found that $C_4$ for 1st peak changes from positive to negative values around $t^* \sim 246 (\nu=0.69)$ and $53 (\nu=0.65)$, respectively.
On the contrary, $C_4$ for 2st peaks changes negative to positive values around $t^* \sim 197 (\nu=0.69)$ and $39 (\nu=0.65)$, respectively.
This is because the geometry of configurations between $i-j$ and $k-l$ pairs.
Those results indicate that $C_4$ in $\nu=0.65$ decays much faster than that of $\nu=0.69$.
In our new method, $C_4(\Delta R, t)$ can resolve how the cluster size distribution changes in time for each density quantitatively.

\subsection{Generalized Order Parameter}

\begin{figure*}
\begin{minipage}{0.47\hsize}
\begin{center}
\includegraphics[scale=0.54]{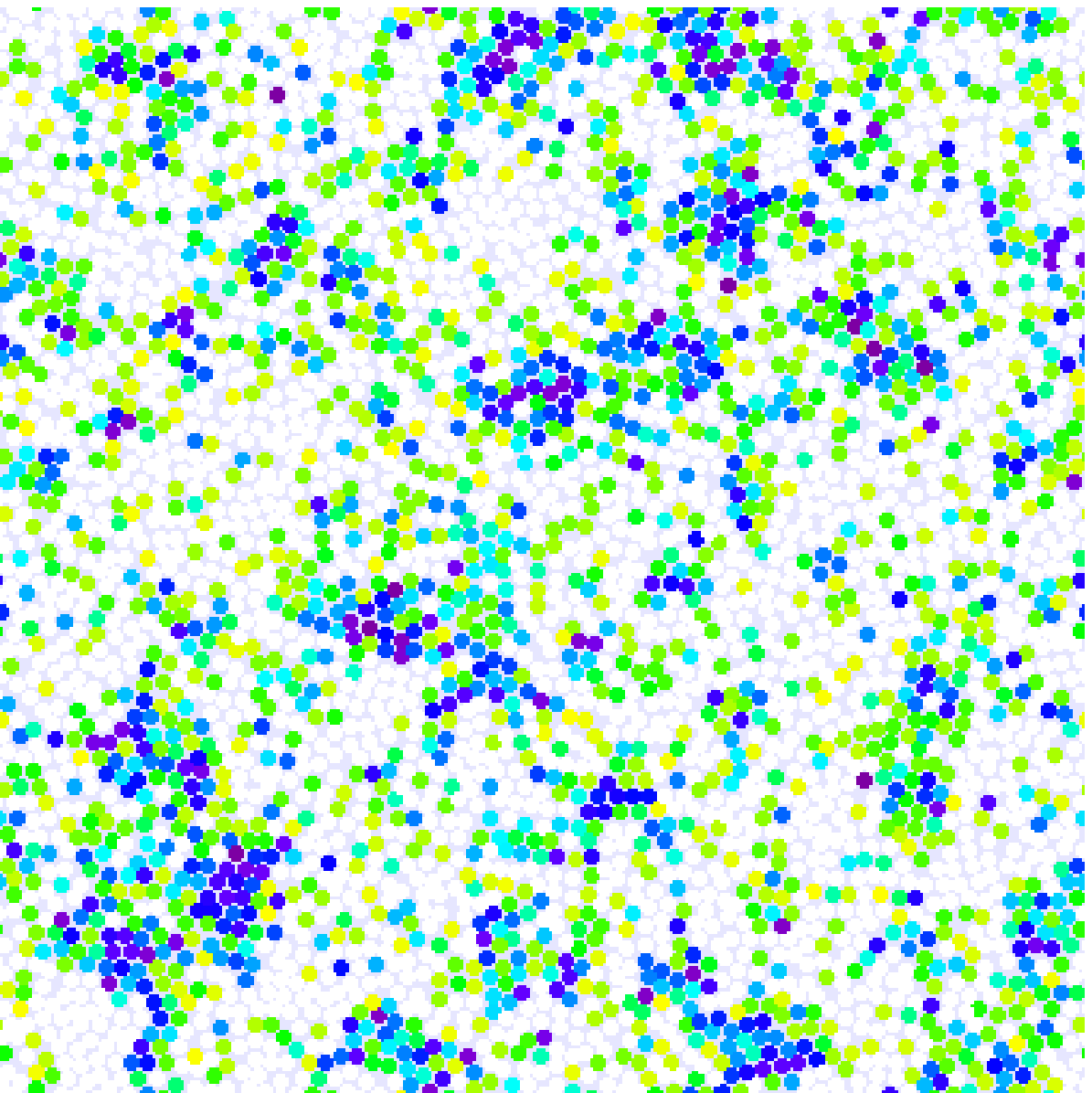}
\end{center}
\begin{center}
0.5 \includegraphics[scale=0.13]{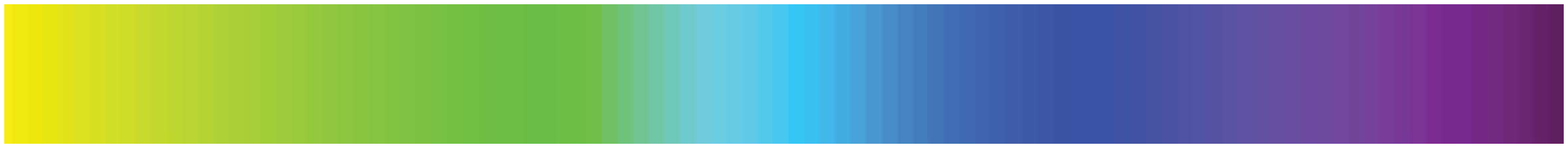} 1
\end{center}
\end{minipage}
\begin{minipage}{0.47\hsize}
\begin{center}
\includegraphics[scale=0.54]{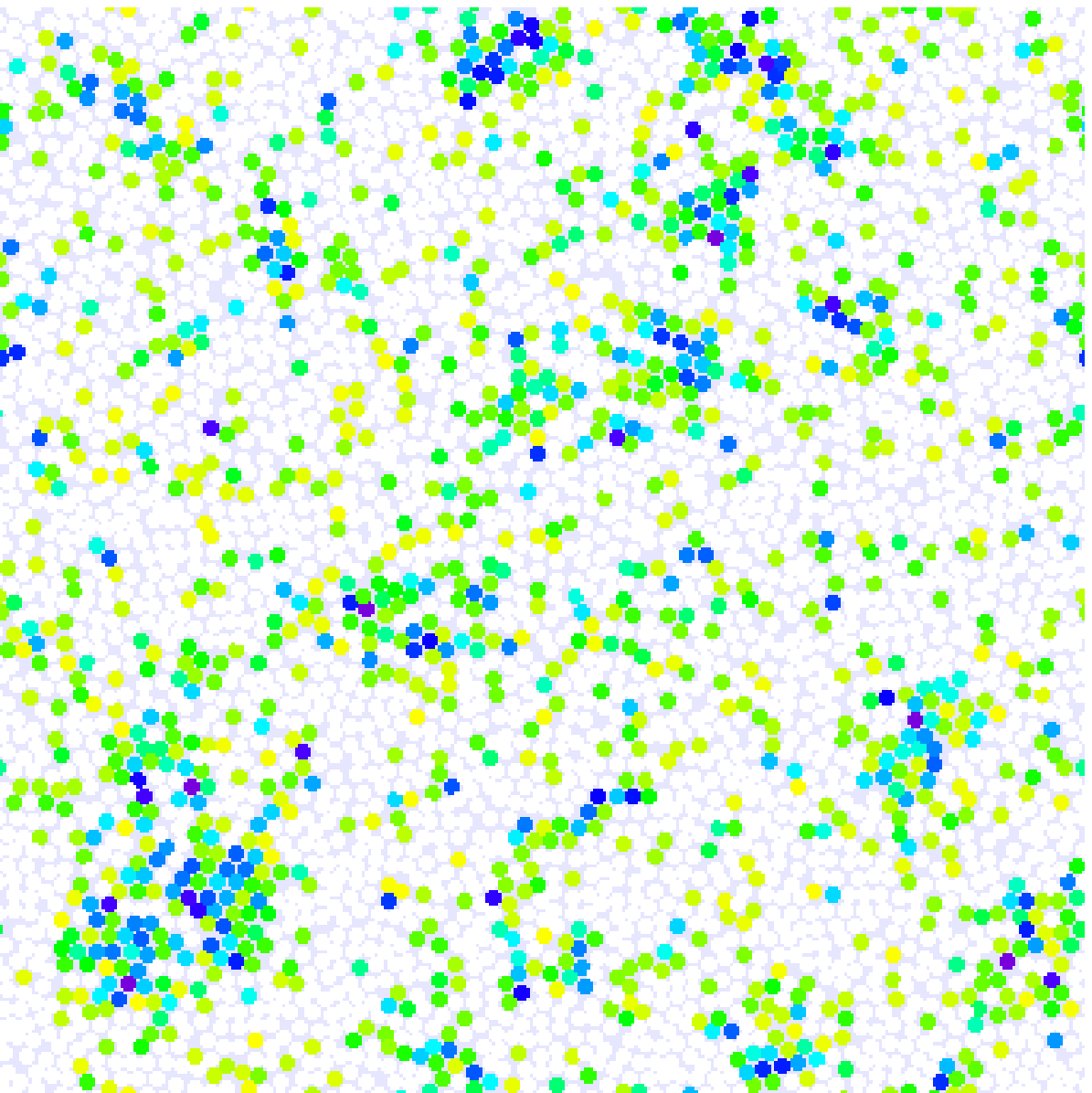}
\end{center}
\begin{center}
0.5 \includegraphics[scale=0.13]{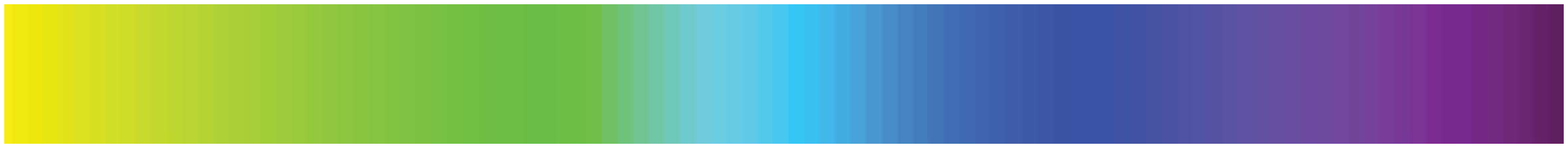} 1
\end{center}
\end{minipage}
\caption{The spatial distribution of generalized order parameter $\Phi_s^i$ (for $s=12$ (left) and $s=18$ (right)) in $4096$ particles system at a given time at $\nu=0.69$.
The darker the region, the closer $\Phi_s^i$ is to unity.
For $s=6$, see Fig.~\ref{fig:phi6_snapshot}(right).
Note that GOPs of higher order N.N. for a given central particles may occasionally have a higher correlation than that of a lower order N.N. because the core particles may not be as highly ordered.
}
\label{fig:phis_snapshot}
\end{figure*}

Tables~\ref{tbl:GOP_typeI} shows the results for the generalized order parameters (GOP), $\Phi_s^i=\sqrt{{\phi_s^i}^* \phi_s^i}$, for $s=6, 12, 18$ (i.e., 1st -3rd N.N.) as described in Sec.II.D for each packing fraction $\nu=0.57$, $0.65$, and $0.69$.
In Table~\ref{tbl:GOP_typeI}, the GOPs are calculated based on the fixed reference axis (i.e. $x$-axis, ${\bf r}_{J'I}=(1,0)$).
We note that all GOPs decrease for higher neighbors for each packing fraction and increase for the higher packing fraction.
Figure~\ref{fig:phis_snapshot} shows the spatial distribution of GOP $\Phi_s^i$ ($s=12$ (left) and $s=18$ (right)) for $4096$ particles system at a given time at the packing fractions $\nu=0.69$.
Comparing with Fig.~\ref{fig:phi6_snapshot}, the darker the region in Fig.~\ref{fig:phi6_snapshot} gradually decreases for 2nd and 3rd N.N.
This can be useful to determine the crystal size quantitatively and to further characterizing the transient clusters.

\begin{table}
\begin{center}
\begin{tabular}{cccc} \hline \hline
\quad $\nu$ \quad & 0.57 & 0.65 & 0.69 \\ \hline
 $\langle \Phi_6^i\rangle$ &    0.487 & 0.556 & 0.645 \\ 
 $\langle \Phi_{12}^i\rangle$ & 0.369 & 0.416 & 0.496 \\ 
 $\langle \Phi_{18}^i\rangle$ & 0.300 & 0.341 & 0.402 \\ 
\hline \hline
\end{tabular}
\end{center}
\caption{
The generalized order parameter, $\Phi_s^i$, at various packing fractions.}
\label{tbl:GOP_typeI}
\end{table}

\subsection{Autocorrelation Function Based on Generalized Order Parameter}

The autocorrelation functions based on the generalized order parameter for each particle $i$ is decomposed into real and imaginary parts,

\begin{equation}
\phi_s^i(t) = {\rm Re}(\phi_s^i (t))+{\rm i}{\rm Im}(\phi_s^i (t)),
\end{equation}

\noindent
leading to the following autocorrelations,

\begin{eqnarray}
C_{\rm GOP}(t) & = & \langle {\phi^*_s}^i(t)\phi_s^i(0)\rangle \\
& = & \langle {\rm Re}(\phi_s^i(t)) {\rm Re}(\phi_s^i(0)) + {\rm Im}(\phi_s^i(t)) {\rm Im}(\phi_s^i(0)) \rangle, 
\label{eqn:acf_typeI}
\end{eqnarray}

The decay of this autocorrelation function $C_{\rm GOP}(t)$ for $s=6,12,18$ at $\nu=0.69$ and $0.65$ normalized by $C^{*}_{\rm GOP}(t)=C_{\rm GOP}(t)/\langle\Phi_s^i\rangle^2$ is given in Table \ref{tbl:Relaxation}.
The decay of the shear stress is directly related to the decay of $C$ (see, Fig.~\ref{fig:2d_c2_cg} and Table \ref{tbl:Relaxation}), and its life time to the dissolution of the cluster, i.e., when its core, $C_2$, melts ~\cite{isobe_2009,isobe_2010}.
A movie of a transient crystal nuclei formation shows that the growing process involves the increasing order of the neighboring particles to as large as 3rd neighbors followed by the dissolving process till the nearest neighbors are no longer ordered.
The overall average of all nucleation processes has a time scale given in Table \ref{tbl:Relaxation} at $\nu=0.69$ of $40, 20, 15$ for the 1st, 2nd, and 3rd N.N., respectively.



\section{Summary and Discussion}

\begin{figure}
\begin{center}
\includegraphics[scale=0.45]{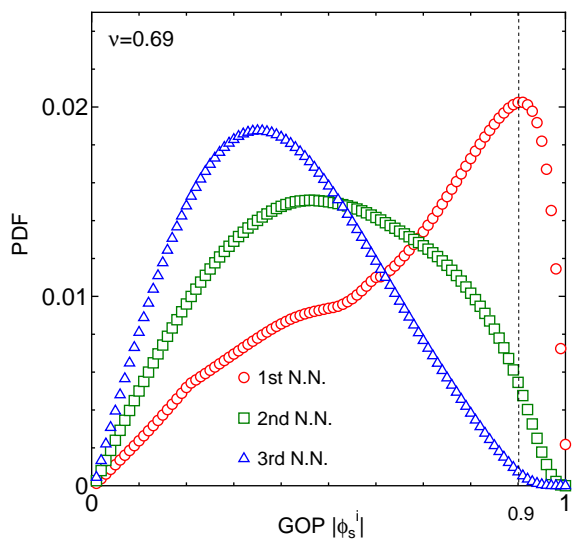}
\end{center}
\caption{
The probability distribution functions of $\Phi_s^i$ at $\nu=0.69$ for further neighbors.
}
\label{fig:phi_hist_nn}
\end{figure}

\begin{figure*}
\begin{minipage}{0.47\hsize}
\begin{center}
\includegraphics[scale=0.45]{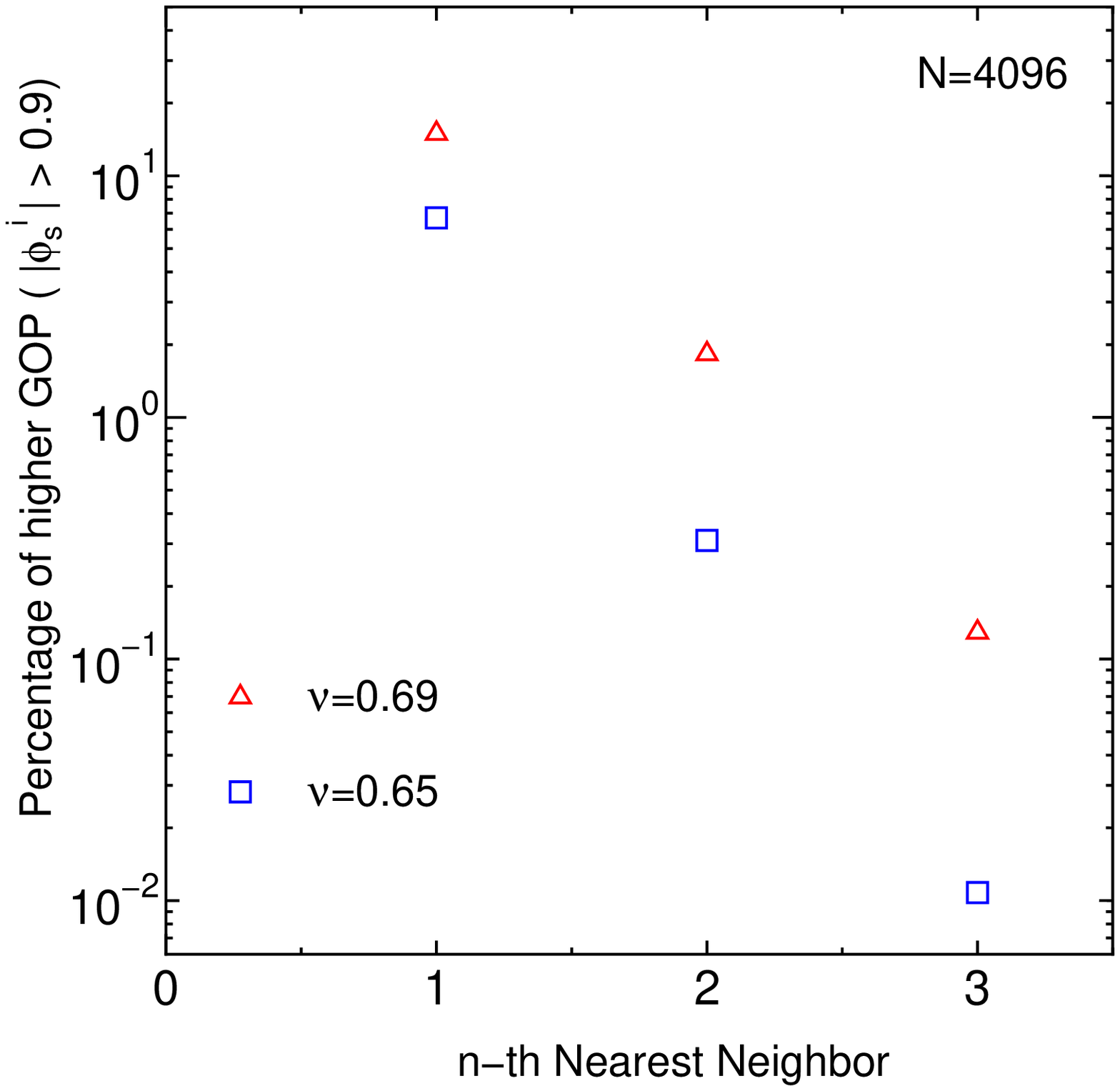}
\end{center}
\end{minipage}
\begin{minipage}{0.47\hsize}
\begin{center}
\includegraphics[scale=0.58]{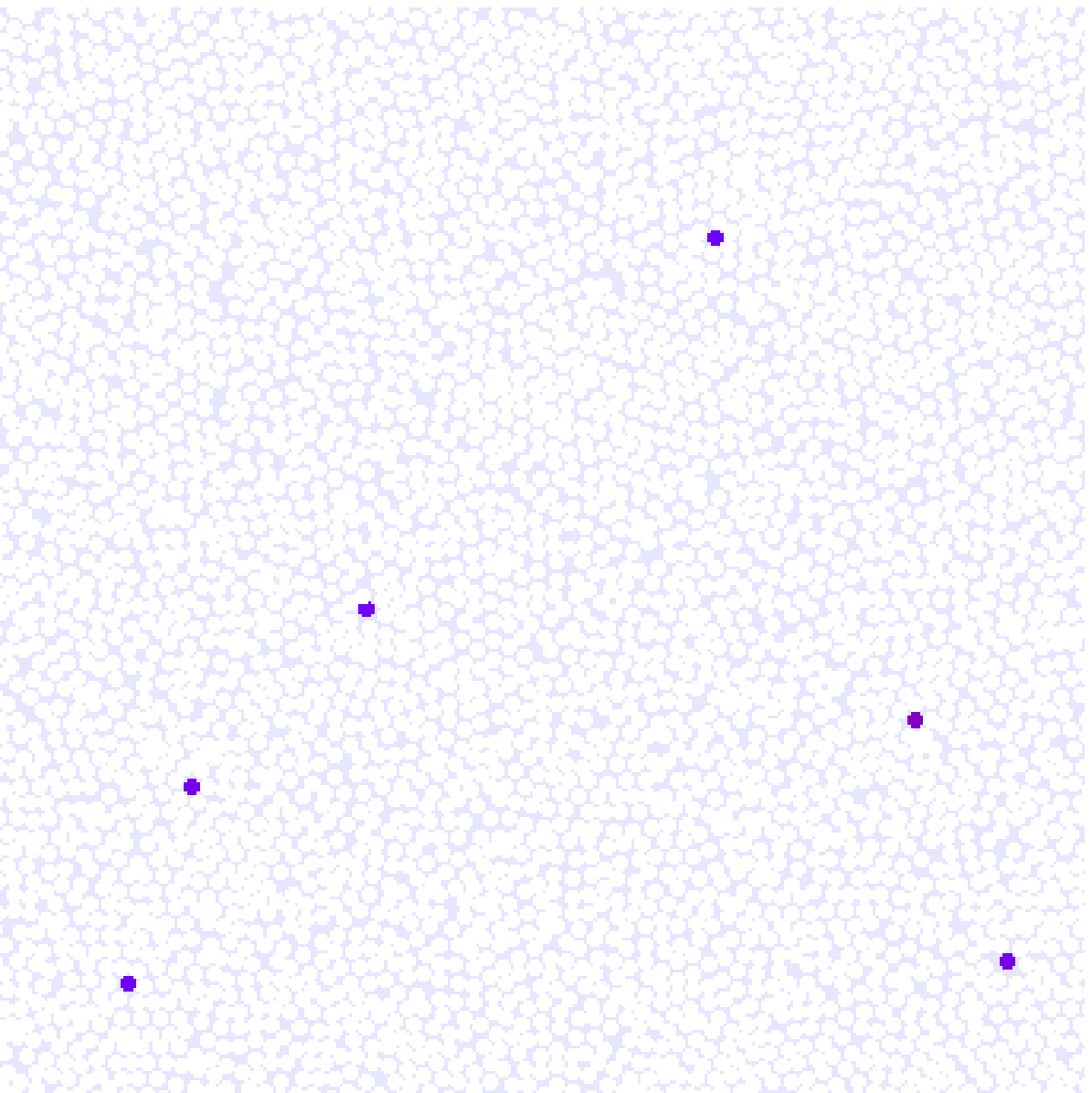}
\end{center}
\end{minipage}
\caption{
(left) The percentage of GOP with $\Phi_s^i > 0.9$ in terms of  $n-th$ N.N. for $\nu=0.65, 0.69$ are shown.
(right) The distribution of GOP for the central particle of a cluster is plotted using threshold $\Psi_s^i > 0.9$ for 3rd N.N. in the same configuration as Fig. 9. }
\label{fig:per_high_phi_nn}
\end{figure*}

In this paper, a method for analysing higher order parameter of the liquid state is developed, especially to investigate transient crystals in dense liquid systems.
Instead of calculating orientational factors and their decomposition based on collision event as previously~\cite{isobe_2009,isobe_2010}, we developed a more efficient methodology by introducing neighbor shells which coarse grain collision events within these neighbor shells as candidates of further collisions during short times.
We confirmed that the same results and relaxation time are obtained as the previously.
We also demonstrate that this improvement permits calculating higher order parameters of orientational contributions such as the distance dependence of the autocorrelation function of the quadruplet contributions $C_4(\Delta R,t)$, providing information on the size and number of the transient crystals and their life time.
The size and number distribution can be investigated not only for 1st N.N. but also for further neighbors, as shown in Fig.~\ref{fig:phi_hist_nn}.
If we somewhat arbitrary recognize particles with $\Phi_s^i > 0.9$ as the center of crystal-like structures as before~\cite{isobe_2010}, it is possible to estimate the crystalline fraction for higher neighbors, as shown in Fig.~\ref{fig:per_high_phi_nn}.

At $\nu=0.69$, we found that $\sim 15 \%$ of the particles in the system have crystal-like nearest neighbors, $\sim 1.8\%$ 2nd N.N. and $\sim 0.13\%$ 3rd N.N.
This means that there are $\sim 5$ (i.e., $= 4096 \times 0.0013$) crystal clusters of the size of 3rd N.N. shell or the size of $\sim 36$ particles in the system.
The typical snapshot of the spatial distribution confirmed the above estimation of size and cluster number. (See, the right of Fig.~\ref{fig:per_high_phi_nn}).
Rarely have the central particles and some of the next nearest neighbor order beyond $\Psi_s^i > 0.9$ at the same time because such 3rd order N.N. are rare events.

The lifetime of these clusters given in the unit of $t_0$ (mean free time) are determined from the autocorrelations for $C_{total}(t)$, $C_2(t)$, $C_4(\Delta R, t)$ and $C_{\rm GOP}(t)$ that are of the stretched exponential form.
The relaxation time is defined as the time when the auto-correlation function decays to $1/e$ of its initial value.
\begin{table}
\begin{center}
\begin{tabular}{cccccccc} \hline \hline
$\nu$ & $\tau (C_{total})$ & $\tau ({C_2})$ & $\tau_1 (C_4)$ & $\tau_2 (C_4)$ & $\tau_{\rm GOP}^1$ & $\tau_{\rm GOP}^2$ & $\tau_{\rm GOP}^3$ \\ \hline
$0.69$ & $12$ & $133$ & $78$  & $66$ & $40$ & $20$ & $15$ \\
$0.65$ & $6$ & $41$ & $21$ & $21$ & $9$ & $7$ & $6$\\
\hline \hline
\end{tabular}
\end{center}
\caption{
Relaxation time $\tau$ for $C_{total}$, $C_2$, and $\tau_k$ of $C(\Delta R, t)$ for 1st ($k=1$) and 2nd ($k=2$) peaks, and $\tau_{\rm GOP}^n$ for $n-$th N.N. at $\nu=0.69$ and $0.65$. All time are in the unit of $t_0$ (mean free time).
}
\label{tbl:Relaxation}
\end{table}
In Table~\ref{tbl:Relaxation}, the relaxation time $\tau$ for $C_{total}$, $C_2$, and $\tau_k$ of $C(\Delta R, t)$ for $k$-th peaks, and $\tau_{\rm GOP}^s$ for $n$-th N.N. at $\nu=0.69$ and $0.65$ are summarized.
As expected, the relaxation time increase when the packing fraction increases and decreases for the higher neighbor shells.
In comparing the relaxation time for $C_2$, $C_4$ for the 1st peak and GOP for 
1st N.N., account must be taken of the number of particles involved, namely 2, 4 and 6 particles respectively.
The larger the numbers, the quicker their order is destroyed.
$C_4(\Delta R, t)$ for the 1st peak loses orientational order faster than $C_2$ and $\tau_{\rm GOP}^1$ even faster.
Interestingly the relaxation curves for $C_4$ at $\nu=0.65$ are almost the same for the first and second peaks although the amplitudes are different,
while $C_4(\Delta R, t)$ for the 1st peak at $\nu=0.69$ has a somewhat slower decay than that for the 2nd peak.
The rapid increase in relaxation time of clusters with density as well their number and size is closely related to the rapid increase in viscosity near freezing.
If we cool the system or compress much faster than the relaxation time estimated by the methods presented here, we can determine the condition under which glass might form.



\newpage
\begin{acknowledgments}
M.I. is grateful to Profs.~W. Kob, L.~Berthier, and H.~Mori for helpful discussion.
This study was supported by Grant-in-Aid for Scientific Research from the Ministry of Education, Culture, Sports, Science and Technology No. 23740293.
This paper is financially supported by Nitto Foundation.
Part of the computations was performed using the facilities of the Supercomputer Center, ISSP, Univ. of Tokyo, and RCCS, Okazaki, Japan.
This work is performed with the support and under the anspices of the NIFS Collaboration Research program (NIFS11KNTS010, NIFS11KNSS020).
\end{acknowledgments}


\end{document}